\documentclass[10pt,english]{article}
\usepackage[T1]{fontenc}
\usepackage[latin1]{inputenc}
\usepackage{geometry}
\usepackage{graphicx}

\makeatletter

\providecommand{\tabularnewline}{\\}

\usepackage{graphicx}

\usepackage{babel}
\makeatother
\begin{document}

\title{Relativistic coupled-cluster calculation of core ionization potential
using the Fock space eigenvalue independent partitioning technique }

\author{Chiranjib Sur and Rajat K Chaudhuri\\
 \emph{NAPP Group, Indian Institute of Astrophysics, Bangalore - 560
034, INDIA}}

\maketitle
\begin{abstract}
In this paper we have applied the cluster-expansion ansatz for the
wave operator $\Omega$ which incorporates the orbital relaxation
and correlation effects in an efficient manner. We have used both
ordinary and normal ordered cluster operator ($\Omega$) to compute
the ionization potential for different states of Ne, Ar, and Kr. The
result in our calculation clearly shows how the relaxation and correlation
effects play the role in determining the ionization potential, one
of the interesting aspects in theoretical spectroscopy.

~

\textbf{PACS number(s).} : 31.10.+z, 31.15.Ar, 32.10.Hq, 31.15.Dv 
\end{abstract}

\section{\label{intro}Introduction}

In spectroscopic studies ionization potential (IP), electron affinity
(EA), excitation energy (EE) and double IP (DIP) are important in
many aspects. For example, in all kind of electron spectroscopy, including
photo-electron and Auger the IP, DIPs always play major roles. Specially,
the knowledge of core IP provides insights into the choice and tunability
of the ionizing beam in all kinds of electron scattering experiments
like $e-2e$, $e-3e$, $\gamma-2e$, double Auger decay etc. These
experiments are not only important to understand the physics but also
to the development of new fields like quantum computing thorough the
idea of entanglement involved in the process. Hence accurate theoretical
determination of IP for core as well as valance electrons may give
some impetus to the physics underlying in those experiments.

Open-shell Coupled Cluster (OSCC) method is used since several years
to compute the IP, EA \cite{rajat-1,spal-1,spal-2} and DIPs \cite{rajat-1,dhiman-1}
- usually employing complete model space (CMS) \cite{cms-1,cms-2}.
In 1980's coupled cluster (CC) methods have been generalized to handle
the incomplete model space (IMS) \cite{spal-1,dhiman-1,kaldor-1,kaldor-2}
in a size-extensive manner. In this paper we have tried to demonstrate
the results arising from the algebraic structure of the expansion
of the wave operator $\Omega=\exp(T+S)$. We have used both the ordinary
and normal ordered cluster expansion of the wave operator $\Omega$.
Normally the operator $T$ corresponds to core-correlation.

Lindgren \cite{lindgren} has shown how the normal ordered cluster
expansion of the wave operator $\Omega$ can include the correlation
effects. Here we will show in a sophisticated way how the relaxation
effects, the many-body effect of different kind can also be taken
into account in the calculation of ionization potential within the
algebraic structure of the semi-normal ordered cluster expansion of
the wave operator $\Omega$. The form of the cluster expansion takes
care the relaxation effect in a compact and efficient manner and it
can be easily demonstrated how the coupling of different many body
effects like relaxation and correlation effects come into the picture
automatically. In an earlier publication Chaudhuri \emph{et. al.}
\cite{rajat-ip1} have explored this techniques and applied to molecules.

In this paper we have calculated ionization potentials (IPs) of different
states of Ne, Ar and Kr. IPs are calculated for core and valance orbitals
of those closed shell atoms using two different approaches. Section
\ref{theory} deals with the theoretical formulation of the two different
methods of calculation using the $\exp(S)$ and $\left\{ \exp(S)\right\} $
theories respectively, where $\left\{ \cdots\right\} $ denotes normal
ordering. Section \ref{results} presents the results obtained from
our calculation and a detailed discussion about the physical aspects
of the two theories. In section \ref{concl} we have made the concluding
remarks and the important findings of our calculations.

\section{\label{theory}Theory}

It is a well established fact that the orbital relaxation effects
are more important than the correlation effects in the core hole ionization
(CHI) process. Since in CHI the relaxation effects are stronger than
the correlation effects, perturbative treatment will not be a efficient
one because in higher orders it will be very cumbersome and that is
the reason to choose the non-perturbative approach like CC.

Let us start from the $N$-electron Dirac Fock (DF) determinant $|\Phi_{0}\rangle$.
The $(N-1)$ ionized determinants $|\Phi_{\alpha}\rangle$ are obtained
from the DF reference state where one electron is ionized from the
orbital $\alpha$. Here the labels $\alpha,\beta\ldots$ denote holes
and $p,q\ldots$ denote particles. According to Thouless\cite{thouless},
the orbital relaxation corrections can be introduced through an exponential
transformation involving one body operator of the form: \begin{equation}
|\Phi_{\alpha}^{{\displaystyle {\displaystyle {\displaystyle {\scriptstyle \mathrm{mod}}}}}}\rangle=\exp(S_{1}^{\alpha})|\Phi_{\alpha}\rangle=\exp(S_{1}^{\alpha})a_{\alpha}|\Phi_{0}\rangle,\label{eq1}\end{equation}
 where \begin{equation}
S_{1}^{\alpha}=\sum_{\begin{array}{c}
{\scriptscriptstyle p\notin\left|\Phi_{\alpha}\right\rangle }\\
{\scriptstyle {\scriptscriptstyle \gamma\in\left|\Phi_{\alpha}\right\rangle }}\end{array}}\left\langle p\right|s^{\alpha}\left|\gamma\right\rangle a_{p}^{\dagger}a_{\gamma},\label{eq2}\end{equation}
 provided \begin{equation}
\langle\Phi_{\alpha}^{\mathrm{mod}}|\Phi_{\alpha}\rangle\ne0.\label{eq3}\end{equation}
 The amplitudes $\langle p|s^{\alpha}|\gamma\rangle$ can be uniquely
determined and orbital relaxation effects may be induced through the
above exponential transformation. Like Eq.(\ref{eq1}), $\left|\Psi\right\rangle $,
the combination of the determinantal states $\left|\Phi_{\alpha}\right\rangle $
also transforms in a similar way and we have the new function having
modified determinant with the associated modified orbital set for
each determinant. We write it as \begin{equation}
|\Phi_{\alpha}^{{\displaystyle {\displaystyle {\displaystyle {\scriptstyle \mathrm{mod}}}}}}\rangle=\sum_{\alpha}\exp(S_{1}^{\alpha})P_{\alpha}|\Psi\rangle=\sum_{\alpha}C_{\alpha}\exp(S_{1}^{\alpha})|\Phi_{\alpha}\rangle,\label{eq4}\end{equation}
 with \begin{equation}
|\Psi\rangle=\sum_{\alpha}C_{\alpha}|\Phi_{\alpha}\rangle,\label{eq5}\end{equation}
 where $C_{\alpha}$'s are the coefficients of $\Psi$ and the $P_{\alpha}$'s
are the projectors onto the $\Phi_{\alpha}$.

\begin{figure}
\begin{center}\includegraphics{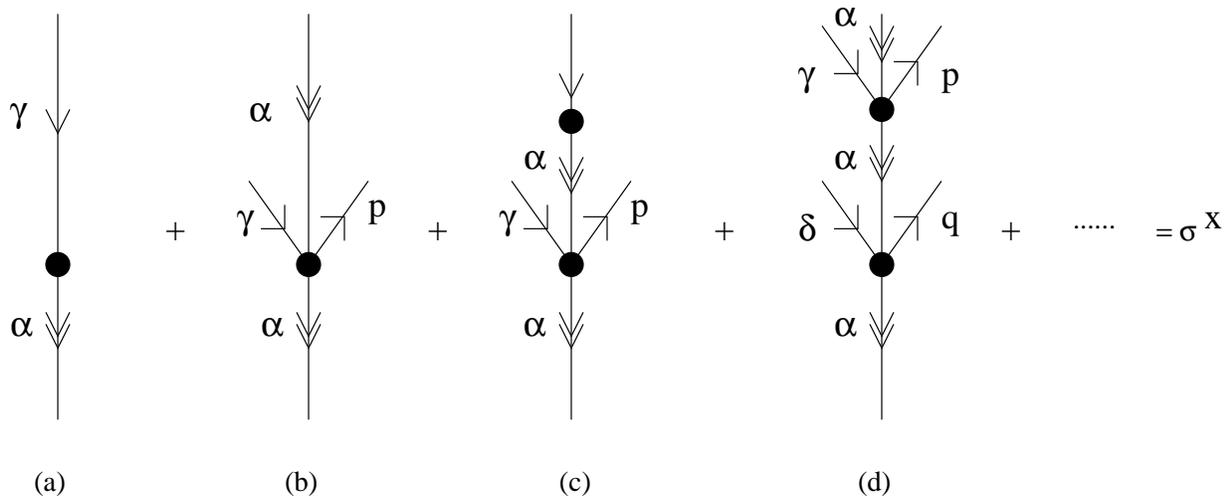}\end{center}

\caption{\label{fig-1} Hugenholtz type diagrams to show the relaxation effects}
\end{figure}

\begin{figure}
\begin{center}\includegraphics{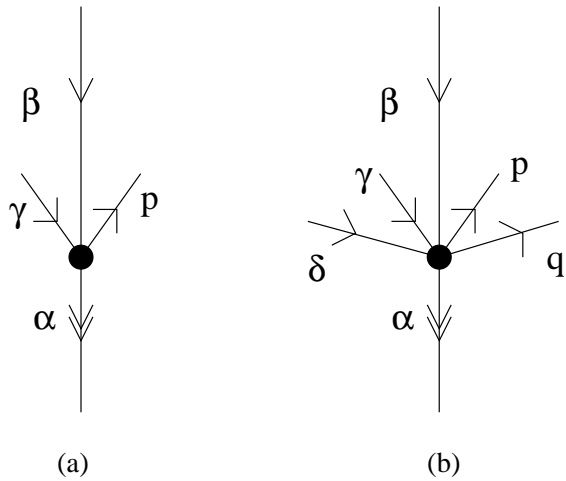}\end{center}

\caption{\label{fig-2}Diagrammatic representation of the cluster operator
$S$}
\end{figure}

Now our aim is to define an expansion which is analogous to Eq.(\ref{eq4}).
To do so, we introduce a valence universal wave operator $\Omega$
\cite{lindgren} for both neutral and the (N-1) electron ionized model
space states, satisfying the Fock-space Bloch equations \cite{cms-1,cms-2,lindgren}\begin{equation}
H\Omega P^{(k)}=\Omega P^{(k)}H_{eff}P^{(k)}.\label{eq6}\end{equation}
 where $P^{(0)}$ and $P^{(1)}$ are the projectors onto $\Phi_{0}$
and $\left\{ \Phi_{\alpha}\right\} $, respectively. We invoke a \emph{normal
ordered} exponential ansatz \cite{lindgren} for $\Omega$ as \begin{equation}
\Omega=\Omega_{c}\Omega_{v}=\exp(T)\left\{ \exp(S)\right\} \label{eq7}\end{equation}
 that satisfies \begin{equation}
|\Psi\rangle=\sum_{\alpha}\Omega C_{\alpha}|\Phi_{\alpha}\rangle.\label{eq8}\end{equation}
 The cluster operators $T$ and $S$ appearing in Eq.(\ref{eq7})
are defined as \begin{equation}
T=\sum_{p,\gamma}\langle p|t_{1}|\gamma\rangle+\sum_{p,q,\gamma,\beta}\langle pq|t_{2}|\gamma\beta\rangle\left\{ a_{p}^{\dagger}a_{q}^{\dagger}a_{\beta}a_{\gamma}\right\} ,\label{eq9}\end{equation}
 and \begin{equation}
S=\sum_{\gamma\ne\alpha}\langle\alpha|s_{1}^{\alpha}|\gamma\rangle+\sum_{p,\gamma,\beta}\langle p\alpha|s_{2}^{\alpha}|\gamma\beta\rangle\left\{ a_{p}^{\dagger}a_{\alpha}^{\dagger}a_{\beta}a_{\gamma}\right\} .\label{eq10}\end{equation}
 Here $\{\cdots\}$ denotes normal ordering with respect to $\Phi_{0}$.
Due to the normal ordering, $S$ acting on $\Phi_{\alpha}$ will give
zero unless the determinantal state has a hole in $\alpha$. For convenience,
we express the cluster operator $S$ in terms of correlation ($S^{c}$)
and relaxation ($S^{r}$) operator as \begin{equation}
S=S^{r}+S^{c}\label{eq11}\end{equation}
 where \begin{equation}
S^{r}=\sum_{\gamma\ne\alpha}\langle\alpha|s_{1}^{\alpha}|\gamma\rangle+\sum_{p,\gamma,\alpha}\langle p\alpha|s_{2}^{\alpha}|\gamma\alpha\rangle\left\{ a_{p}^{\dagger}a_{\alpha}^{\dagger}a_{\alpha}a_{\gamma}\right\} ,\label{eq12}\end{equation}
 and \begin{equation}
S^{c}=\sum_{p,\gamma,\beta\ne\alpha}\langle p\alpha|s_{2}^{\alpha}|\gamma\beta\rangle\left\{ a_{p}^{\dagger}a_{\alpha}^{\dagger}a_{\beta}a_{\gamma}\right\} .\label{eq13}\end{equation}
 Figures (\ref{fig-1}a)-(\ref{fig-1}b) represent the Hugenholtz
diagrams corresponding to the terms given in Eq.(\ref{eq12}) and
figure (\ref{fig-2}b) represents Eq.(\ref{eq13}) for $\beta\neq\alpha$.

Since $\left\{ S\right\} ^{2}$ and higher powers of $S$ contains
more than one valence destruction operator $a_{\alpha}^{\dagger}$
and contraction among $S$ operators is not allowed, Eq.(\ref{eq7})
reduces to \begin{equation}
\Omega=\Omega_{c}\Omega_{v}=\Omega_{c}\left\{ 1+S^{r}+S^{c}\right\} .\label{eq14}\end{equation}
 It is evident from Eq.(\ref{eq14}) that $\left\{ \exp(S)\right\} =\left\{ 1+S\right\} $
can not induce relaxation corrections via Thouless-type exponentiation.

Let us now consider expansion of $\Omega$ where $\Omega$ is an ordinary
exponential ansatz, i.e., \begin{equation}
\Omega=\Omega_{c}\Omega_{v}=\exp(T)\exp(S)\label{eq15}\end{equation}
 Substituting Eq.(\ref{eq11}) in Eq.(\ref{eq15}), we get \begin{equation}
\Omega=\Omega_{c}[1+\left\{ S^{c}\right\} +\left\{ S^{r}\right\} +\frac{1}{2!}\left\{ \overline{S^{r}S^{r}}\right\} +\frac{1}{3!}\left\{ \overline{S^{r}S^{r}S^{r}}\right\} +\frac{1}{2!}\left\{ \overline{S^{r}S^{r}}S^{c}\right\} +\cdots].\label{eq16}\end{equation}
 This expansion looks similar to the normal ordered expansion with
the extra property that the contractions like $\overline{S^{r}S^{r}}$
involving the orbital $\alpha$ is allowed. It can easily shown that
in the above expression (Eq.\ref{eq16}), terms like $\left\{ S^{r}\right\} ^{2}$,
$\left\{ S^{c}\right\} ^{2}$ etc. do not contribute because these
terms contain more than one valence destruction operator $a_{\alpha}^{\dagger}$.
Introducing an operator $\sigma^{r}$ through the relation \begin{equation}
\sigma^{r}=\sum_{n=1}\frac{1}{n!}\left\{ \overline{S^{r}\cdots S^{r}\cdots S^{r}}\right\} ,\label{eq17}\end{equation}
 we can rewrite Eq.(\ref{eq16}) in normal order as \begin{equation}
\Omega=\Omega_{c}\left\{ \exp(S^{c}+\sigma^{r})\right\} =\Omega_{c}\Omega_{v}.\label{eq18}\end{equation}
 Eq.(\ref{eq18}) clearly indicates that the ordinary exponential
ansatz $\exp(S)$ is capable of inducing the orbital relaxation effects
through Thouless-type exponentiation. The relaxation effects arising
from the contraction of terms like $\overline{S^{r}S^{r}}$ are shown
in figure \ref{fig-2} where in (\ref{fig-2}a) $\beta,\delta\neq\alpha$
and in figure (\ref{fig-2}b) $\beta,\gamma,\delta\neq\alpha$.

To compute the energy difference, we insert Eq.(\ref{eq14})/Eq.(\ref{eq18})
into Eq.(\ref{eq6}) and proceed hierarchically from $k=0$ to $k=1$.
For $k=0$, Eq.(\ref{eq6}) reduces to closed shell CC equation for
core-cluster amplitudes $T$. For $k=1$, we premultiply Eq.(\ref{eq6})
by $\Omega_{c}^{-1}(\equiv\exp(-T))$, introduce a connected dressed
operator $\widetilde{H}=\Omega_{c}^{-1}H\Omega_{c}-E_{gr}$ and obtain
the Bloch equation for energy difference \begin{equation}
\overline{H}\Omega_{v}P^{(1)}=\Omega_{v}P^{(1)}\widetilde{H}_{eff}P^{(1)},\label{eq19}\end{equation}
 where \begin{equation}
\widetilde{H}_{eff}=H_{eff}-E_{gr}.\label{eq20}\end{equation}

\section{\label{results}Results and discussions}

\begin{figure}
\begin{center}\includegraphics{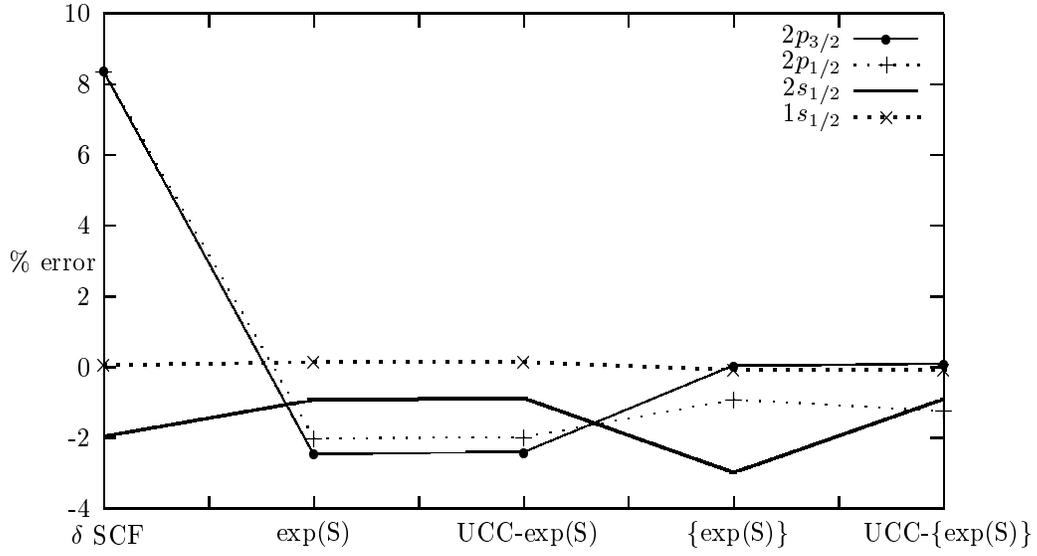}\end{center}

\caption{\label{ip-ne}\% error for different states in computing IP of Ne
using different methods}
\end{figure}

\begin{figure}
\begin{center}\includegraphics{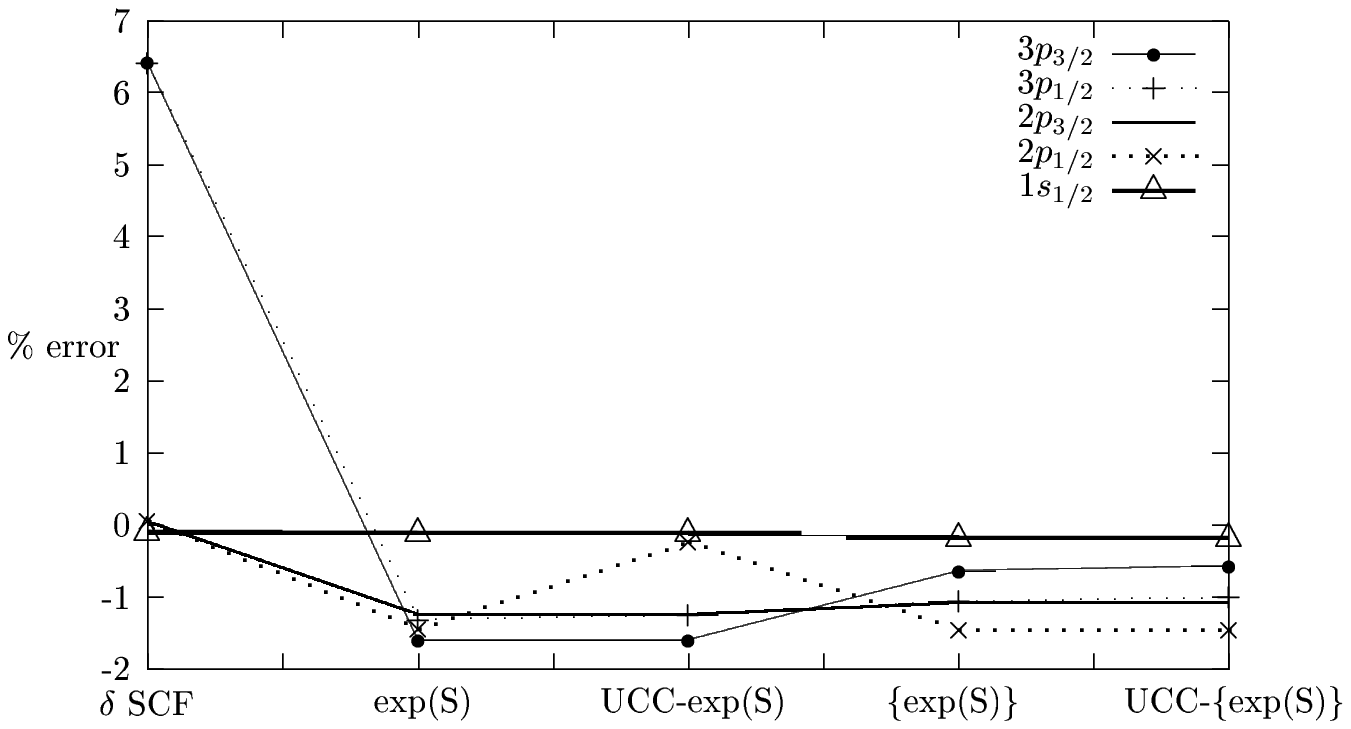}\end{center}

\caption{\label{ip-ar}\% error for different states in computing IP of Ar
using different methods}
\end{figure}

\begin{figure}
\begin{center}\includegraphics{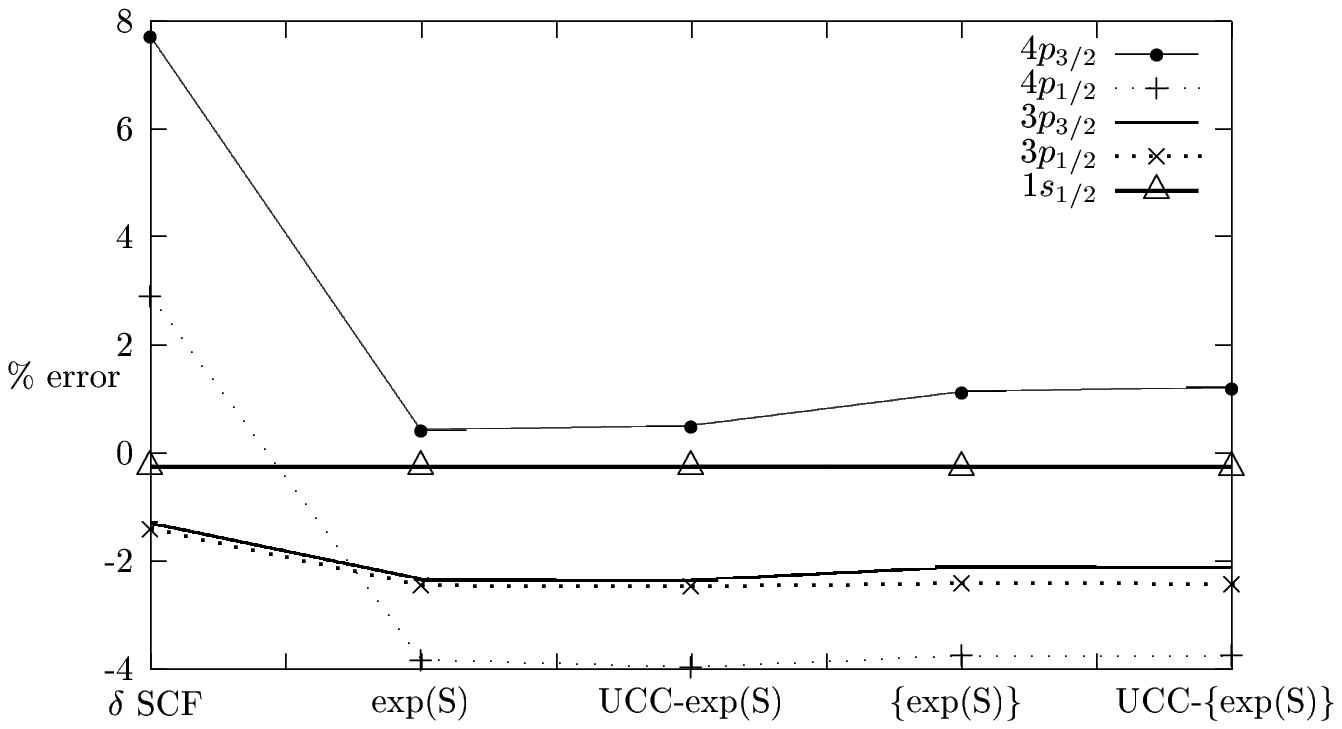}\end{center}

\caption{\label{ip-kr}\% error for different states in computing IP of Kr
using different methods}
\end{figure}

\begin{figure}
\begin{center}\includegraphics{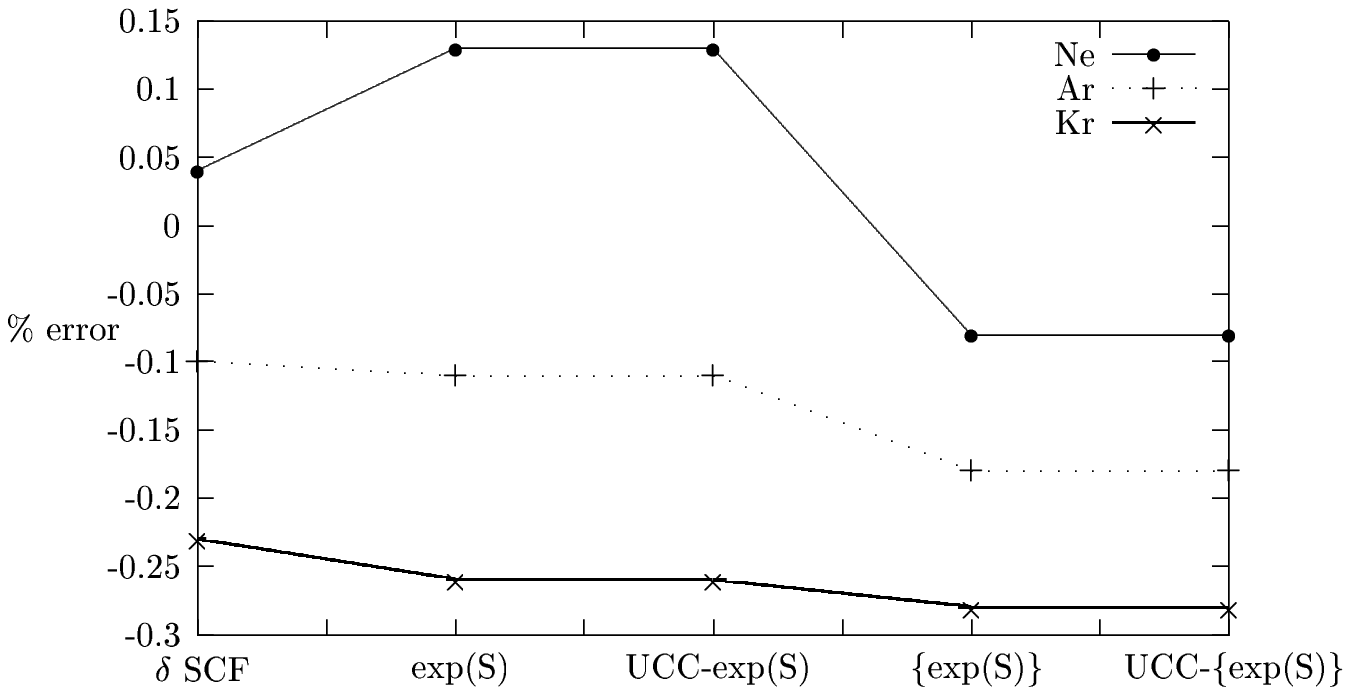}\end{center}

\caption{\label{ip-1s}\% error for different $1s$ states in computing IP
of Ne, Ar and Kr using different methods}
\end{figure}

The results obtained by our calculation of IPs of Ne, Ar and Kr using
two different approaches are presented in table \ref{ip-table}. Figures
\ref{ip-ne}, \ref{ip-ar} and \ref{ip-kr} show the relative error
for all the states of Ne, Ar and Kr respectively, where as the figure
\ref{ip-1s} shows the relative error in IP for $1s$ states explicitly
of the systems studied. From table \ref{ip-table} and figures \ref{ip-ne},
\ref{ip-ar} and \ref{ip-kr} it is clear that the agreement of IPs
obtained by the $\exp(S)$ theory is better than $\left\{ \exp(S)\right\} $
theory for the core states. This nature can be interpreted in terms
of the different many-body effects, namely relaxation and correlation
effects. For deep lying cores, the relaxation effects are stronger
than the correlation effects, which is taken care of by the $\exp(S)$
theory as compared to the $\left\{ \exp(S)\right\} $ theory (Fig.\ref{fig-1},
\ref{fig-2}). Where as for high-lying core orbitals the relaxation
effects are comparatively less than the deep core orbitals like $1s_{1/2}$
and $2s_{1/2}$. And thus the results obtained by the $\left\{ \exp(S)\right\} $
theory performs better than the $\exp(S)$ theory for high lying states
like $3p$ and $4p$. These can be interpreted in a better way if
we look at the Thouless's treatment \cite{thouless}. When the relaxation
is very strong, a perturbative treatment is very cumbersome in higher
orders and also inefficient. The widely used normal ordered cluster
expansion approach is not compact enough to include the relaxation
effects in an adequate manner in the theories to calculate IP, EA,
EE and DIP. To overcome this problem we have explored the generalized
Thouless's treatment and have calculated the results by using the
$\exp(S)$ theory.

\begin{table}

\caption{\label{ip-table}Ionization potentials (in eV) of Ne, Ar and Kr atoms.
Entries within parentheses are from UCC(3).}

\begin{center}\begin{tabular}{lrrrrrr}
\hline 
Atom &
 Ionizing &
 Koopmann&
 $\delta$-SCF &
 $\exp(S)$ Theory &
 $\{\exp(S)\}$ Theory &
 Expt. \cite{webURL}\tabularnewline
&
 Orbital &
&
&
&
&
\tabularnewline
\hline
&
&
&
&
&
&
\tabularnewline
Ne &
 $2p_{3/2}$&
 23.08&
 19.79 &
 22.13 ( 22.12) &
 21.59 ( 21.58) &
 21.6 \tabularnewline
&
 $2p_{1/2}$&
 23.21&
 19.89 &
 22.14 ( 22.13) &
 21.90 ( 21.97) &
 21.7 \tabularnewline
&
 $2s_{1/2}$&
 52.69&
 49.45 &
 48.95 ( 48.93) &
 48.95 ( 48.94) &
 48.5 \tabularnewline
&
 $1s_{1/2}$&
 893.10&
 869.80 &
 869.10 ( 869.06) &
 870.92 ( 870.89) &
 870.2 \tabularnewline
&
&
&
&
&
&
\tabularnewline
Ar &
 $3p_{3/2}$&
 15.99&
 14.69 &
 15.95 ( 15.95) &
 15.80 ( 15.79) &
 15.7 \tabularnewline
&
 $3p_{1/2}$&
 16.20&
 14.88 &
 16.11 ( 16.10) &
 16.07 ( 16.06) &
 15.9 \tabularnewline
&
 $2p_{3/2}$&
 259.79&
 248.26 &
 251.51 ( 251.51) &
 251.05 ( 251.05) &
 248.4 \tabularnewline
&
 $2p_{1/2}$&
 262.09&
 250.48 &
 254.21 ( 251.21) &
 254.25 ( 254.25) &
 250.6 \tabularnewline
&
 $1s_{1/2}$&
 3242.14&
 3209.19 &
 3209.28 (3209.28) &
 3211.64 (3211.64) &
 3205.9 \tabularnewline
&
&
&
&
&
&
\tabularnewline
Kr&
 $4p_{3/2}$&
 13.99&
 13.01&
 14.04 (14.03)&
 13.94 (13.93)&
 14.1\tabularnewline
&
 $4p_{1/2}$&
 14.73&
 13.69&
 14.64 (14.66)&
 14.63 (14.63)&
 14.1\tabularnewline
&
 $3p_{3/2}$&
 226.25&
 217.20&
 219.45 (219.48)&
 218.93 (218.98)&
 214.4\tabularnewline
&
 $3p_{1/2}$&
 234.58&
 225.34&
 227.65 (227.69)&
 227.56 (227.59)&
 222.2\tabularnewline
&
 $1s_{1/2}$&
 14418.61&
 14359.14&
 14362.44 (14362.59)&
 14365.76 (14365.88)&
 14326\tabularnewline
&
&
&
&
&
&
\tabularnewline
\hline
&
&
&
&
&
&
\\
\tabularnewline
\end{tabular}\end{center}
\end{table}

In table \ref{ip-table} the results in the parentheses are obtained
from UCC(3) and it is clear that for these kind of problem the introduction
of the unitary excitation operator does not improve the results much.
The details and the results using UCC(3) \cite{csur-ucc} will be
communicated shortly.

If we minutely observe the results for Kr we can see that the relaxation
and correlation effects are canceled out for almost all the states.
In other words, these two effects are compensated by each other. This
becomes more clear if we see the Koopman's energy for this system
given in table \ref{ip-table}.

Besides the experimental values, IPs from $\delta$-SCF are also listed
in table \ref{ip-table}. The $\delta$-SCF IPs are computed using
GRASP programme \cite{grasp}. In this procedure, the electron binding
energy corresponds to the $\alpha$th orbital is obtained from

\begin{equation}
IP(\alpha)=E(N)-E_{\alpha}(N-1),\label{delta-scf}\end{equation}
 where $E(N)$ and $E_{\alpha}(N-1)$ are the SCF energies of $N$
and $(N-1)$-electron systems respectively. The present calculation
shows that $\delta$-SCF estimates the core IP more accurately than
the valence IP. This is in accordance with our expectation. Note that
the differential electron correlation which is important for valence
IP can not be incorporated through $\delta$-SCF procedure. Since
the orbital relaxation effect can be introduced through $\delta$-SCF
procedure, this yield reasonably accurate core IPs.

\section{\label{concl}Concluding remarks}

In this paper we have presented the calculation of ionization potentials
for core and valance orbitals of some rare gas atoms, namely Ne, Ar
and Kr. We have explored both the $\exp(S)$ and $\left\{ \exp(S)\right\} $
theory to obtain the results. Along with the normal cluster expansion
ansatz we have also obtained the results by using unitary cluster
operator theory (UCC) for both the formalisms.

The close comparison between the results obtained by the two different
approaches are explored and discussed in details. The figures containing
the computational errors for different theories help us to understand
the trends of the obtained results. This calculation clearly demonstrate
the importance of different many-body effects, namely differential
electron correlation and orbital relaxation effects in determining
ionization potential of closed shell atoms.

\begin{verse}
\textbf{Acknowledgments} : One of the author (CS) acknowledge BRNS
for project no. 2002/37/12/BRNS. The computation was carried out on
our group's Xeon PC cluster. 
\end{verse}


\begin{thebibliography}{10}
\bibitem{rajat-1}R. Chaudhuri, S. Guha, D. Sinha and D. Mukherjee, Lec. Notes in Chemistry,
\textbf{52}, ed. U. Kaldor, Springer Verlag, Berlin, 1989. 
\bibitem{spal-1}S. Pal, M. Rittby, R. J. Bartlett, D. Sinha and D. Mukherjee, J. Chem.
Phys \textbf{88}, 4357 (1988) 
\bibitem{spal-2}S. Pal, M. Rittby, R. J. Bartlett, D. Sinha and D. Mukherjee, Chem.
Phys. Lett. \textbf{137}, 273 (1987). 
\bibitem{dhiman-1}D. Sinha, S. K. Mukhopadhyay and D. Mukherjee, Chem. Phys. Lett. \textbf{125},
213 (1986). 
\bibitem{cms-1}D. Mukherjee and S. Pal, Advn. Quantum. Chem. \textbf{20}, 292 (1989)
and the references therein. 
\bibitem{cms-2}I. Lindgren and D. Mukherjee, Phys. Rep. \textbf{151}, 93 (1987) and
the references therein. 
\bibitem{kaldor-1}U. Kaldor, In. J. Quan. Chem. Symp. \textbf{20}, 445 (1986). 
\bibitem{kaldor-2}U. Kaldor, J. Chem. Phys. \textbf{87}, 467 (1987). 
\bibitem{lindgren}I. Lindgren, Int. J. Quan. Chem. \textbf{S12}, 33 (1978). 
\bibitem{rajat-ip1}D. Mukhopadhyay, R. Chaudhuri and D. Mukherjee, Chem. Phys. Lett.
\textbf{172}, 515 (1990). 
\bibitem{thouless}D. J. Thouless, Nucl. Phys. \textbf{21}, 225 (1960). 
\bibitem{kutzelnigg-1}W. Kutzelnigg, J. Chem. Phys \textbf{77}, 3081 (1982). 
\bibitem{webURL}http://www.webelements.com 
\bibitem{csur-ucc}C. Sur, R. K. Chaudhuri, B. K. Sahoo, B. P. Das and D. Mukherjee,
(To be communicated) and the references therein. 
\bibitem{grasp}K. G. Dyall \emph{et al}, Comp. Phys. Comm. \textbf{55}, 425 (1989).\end{thebibliography}
\end{document}